\documentclass{PoS}
\newcommand{\KK}{${\cal KK}$}
\newcommand{\Meu}{\EuScript{M}}

\def\Order#1{${\cal O}(#1)$}
\usepackage{amsmath,amssymb,bbm,color}
\usepackage{graphicx}
\usepackage{epsfig}
\usepackage{euscript}
\usepackage{pslatex}
\usepackage{fancybox}
\usepackage{enumerate}
\usepackage{widetext}
\usepackage{subfigure}
\def\st{\hbox{}} 
\title{A Precision Event Generator for EW Corrections in Hadron Scattering: {\KK}MC-hh}

\ShortTitle{A Precision Event Generator}

\author{\speaker{B.F.L. Ward}\thanks{On Research Leave from Baylor University, Waco, TX, USA during 01/04/18  to 07/31/18  at Werner-Heisenberg-Institut, Max-Planck-Institut fuer Physik, Foehringer Ring 6, 80805 Muenchen, Germany}\\
        Baylor University, Waco, TX, USA\\
        E-mail: \email{BFL\_Ward@baylor.edu}}

\author{S. Jadach\\
        Institute of Nuclear Physics, Polish Academy of Sciences,
        Krakow, Poland\\
        E-mail: \email{Stanislaw.Jadach@cern.ch}}
        
 \author{Z. Was\\
        Institute of Nuclear Physics, Polish Academy of Sciences,
        Krakow, Poland\\
        E-mail: \email{Z.Was@cern.ch}}   
        
 \author{S.A. Yost\\
        The Citadel, Charleston, SC, USA\\
        E-mail: \email{YostS1@citadel.edu}}    

\abstract{{\KK}MC-hh is a precision event-generator for Z production and decay in hadronic collisions, which
applies amplitude-level resummation to both initial and final state photon radiation, including
perturbative residuals exact through ${\cal O}(\alpha^2L)$, together with exact ${\cal O}(\alpha)$ EW matrix element corrections. We present some comparisons to other programs and results showing the effect of multi-photon radiation for cuts motivated by a recent ATLAS W mass analysis. We also show preliminary untuned comparisons of the electroweak corrections of {\KK}MC-hh to those of HORACE, which includes exact ${\cal O}(\alpha)$ corrections with resummed final-state photon radiation.}
\centerline {\bf BU-HEPP-18-09, Nov., 2018}
\FullConference{ICHEP2018, 39th International Conference on High Energy Physics\\
		5-11 July 2018\\
		Seoul, S. Korea}

\begin{document}

The MC \KK{MC}-hh~\cite{kkmchh} is a precision event generator which incorporates a parton shower in the event generator \KK MC4.22~\cite{kkmc422}. The latter event generator is the extension of the event generator \KK MC 4.13~\cite{kkmc413}, which provided exact ${\cal O}(\alpha^2L)$ coherent exclusive exponentiation (CEEX)~\cite{ceex} EW corrections to the processes $e^+e^-\rightarrow f\bar{f},$ to include $q \bar{q}, \; \mu\bar{\mu},\; \tau\bar{\tau}\; \text{and}\; \nu_\ell\bar{\nu}_\ell,$ initial states, where $ \ell = e,\;\mu,\;\tau.$ The built-in shower in \KK{MC}-hh, whose implementation is described in Ref.~\cite{kkmchh}, is that from Herwig6.5~\cite{hwg} but an external shower can be used via the LHE format~\cite{lhe}; the respective PDF interface uses LHAPDF~\cite{lhapdf}. In what follows, we present some results for cuts motivated by the ATLAS W mass analysis~\cite{atlasmw-17} and some comparisons to other programs.\par
The discussion proceeds as follows. We first give the elements of the CEEX~\cite{ceex} realization of exact amplitude-based resummation theory and the specific approach we use to implement concomitant exact NLO EW corrections. This is followed by results based on cuts motivated by the analysis of the W mass in Ref.~\cite{atlasmw-17}, wherein and where-after we also make comparisons with other calculations.\par
The detailed theory of CEEX  resummation is given in Ref.~\cite{ceex}. When we apply this formalism for the process
$q\bar{q}\rightarrow \ell\bar{\ell}+n\gamma, \; q=u,d,s,c,b,t,\ell=e,\mu,\tau,\nu_e,\nu_\mu,\nu_\tau$, the cross section has the form 
\begin{equation}
\sigma =\frac{1}{\text{flux}}\sum_{n=0}^{\infty}\int d\text{LIPS}_{n+2}\; \rho_{\text{CEEX}}^{(n)}(\{p\},\{k\}),
\label{eqn-hw2.1-1}
\end{equation}
where
\begin{equation}
\rho_{\text{CEEX}}^{(n)}(\{p\},\{k\})=\frac{1}{n!}e^{Y(\Omega;\{p\})}\bar{\Theta}(\Omega)\frac{1}{4}\sum_{\text{helicities}\;{\{\lambda\},\{\mu\}}}
\left|\Meu\left(\st^{\{p\}}_{\{\lambda\}}\st^{\{k\}}_{\{\mu\}}\right)\right|^2.
\label{eqn-hw2.1-2}
\end{equation}
Here $\text{LIPS}_{n+2}$ denotes Lorentz-invariant phase-space for $n+2$ particles. Incoming and outgoing fermion momenta are abbreviated as $\{p\}$ and the $n$ photon momenta are denoted by $\{k\}$. The functions $Y(\Omega;\{p\})\;\text{and}\;\bar\Theta(\Omega;k) $ as well as the CEEX amplitudes $\{\Meu\}$ are defined in Refs.~\cite{kkmc422,kkmc413,ceex}.
\KK{MC}-hh uses the DIZET 6.2 library to realize exact ${\cal O}(\alpha)$ EW corrections as described in Ref.~\cite{ceex}. The procedure for combining (\ref{eqn-hw2.1-1})  with a
shower is given in Ref.~\cite{kkmchh}. At this point, we would like to stress the following. The hard photon residuals in the amplitudes $\{\Meu\}$ generate a hard cross section in the context of the standard factorization theorem formula for the Drell-Yan process,
\begin{equation}
\sigma_{\text{DY}}=\int dx_1dx_2\sum_i f_i(x_1)f_{\bar{i}}(x_2)\sigma_{\text{DY},i\bar{i}}(Q^2)\delta(Q^2-x_1x_2s),
\label{eqn-hw2.1-3}
\end{equation}
where the subprocess for the $i$-th $q\bar{q}$ annihilation with $\hat{s}=Q^2$ when the pp cms energy squared is $s$
is given in a conventional notation for parton densities $\{f_j\}$. {\KK}MC-hh realizes multiple gluon radiation and the attendant hadronization for the concomitant shower via backward evolution~\cite{sjos-sh} for the densities as specified in (\ref{eqn-hw2.1-3}). As shown in Ref.~\cite{sjos2} the soft exponentiated QED radiation in (\ref{eqn-hw2.1-1}) is unaffected by
phase space competition with the gluons in the shower and, by the space-time structure of the factorization theorem, the gluons in the shower do not compete with the hard cross section radiation -- quanta on different time scales do not compete: shower gluons compete with shower photons, which we do not have in \KK{MC}-hh, and hard cross section radiation gluons compete with hard cross section radiation photons but shower quanta do not compete with hard cross section radiation quanta, by microscopic causality. 
We turn now to application of \KK{MC}-hh to observables used in the ATLAS $M_W$ measurement.\par
\begin{figure}[h]
\begin{center}
\includegraphics[width=140mm]{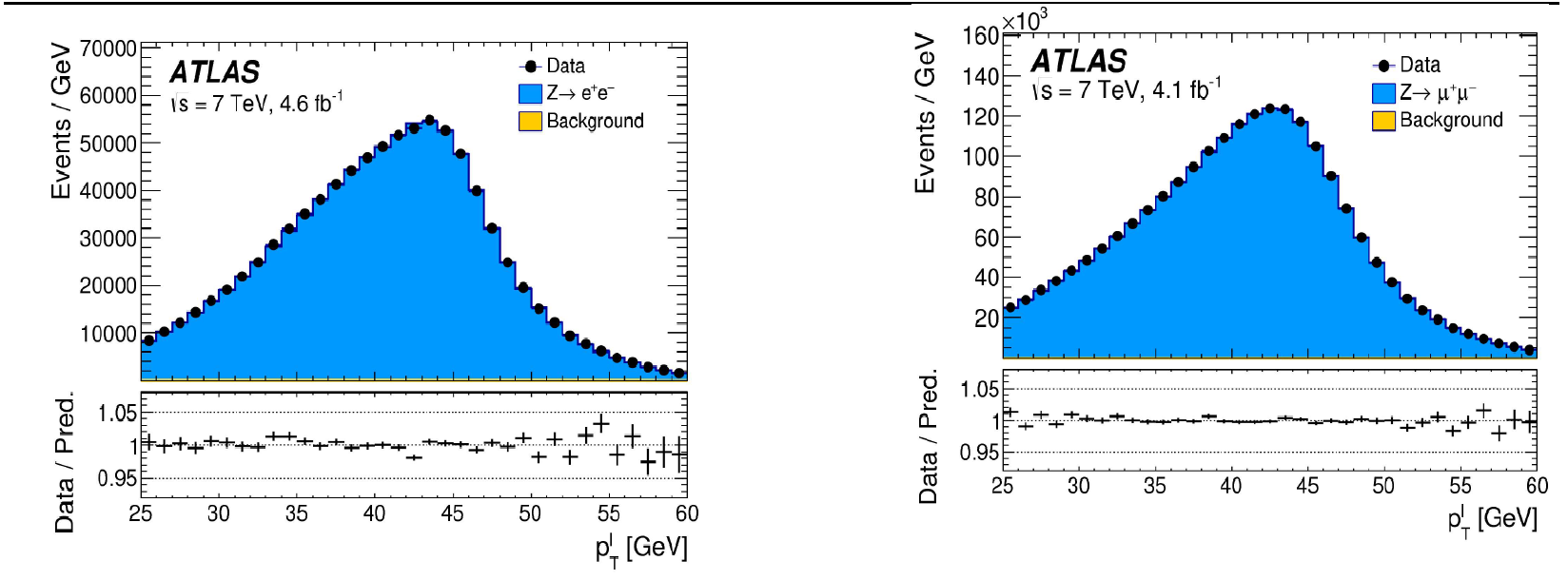}
\end{center}
\caption{\baselineskip=11pt     ATLAS electron and muon $p^{\ell}_T$ spectra from Fig. 15 in Ref.~\cite{atlasmw-17}.}
\label{fig1}
\end{figure}

\begin{figure}[h]
\begin{center}
\setlength{\unitlength}{1in}
\begin{picture}(6,2.4)(0,0)
\put(0,0.2){\includegraphics[width=3in]{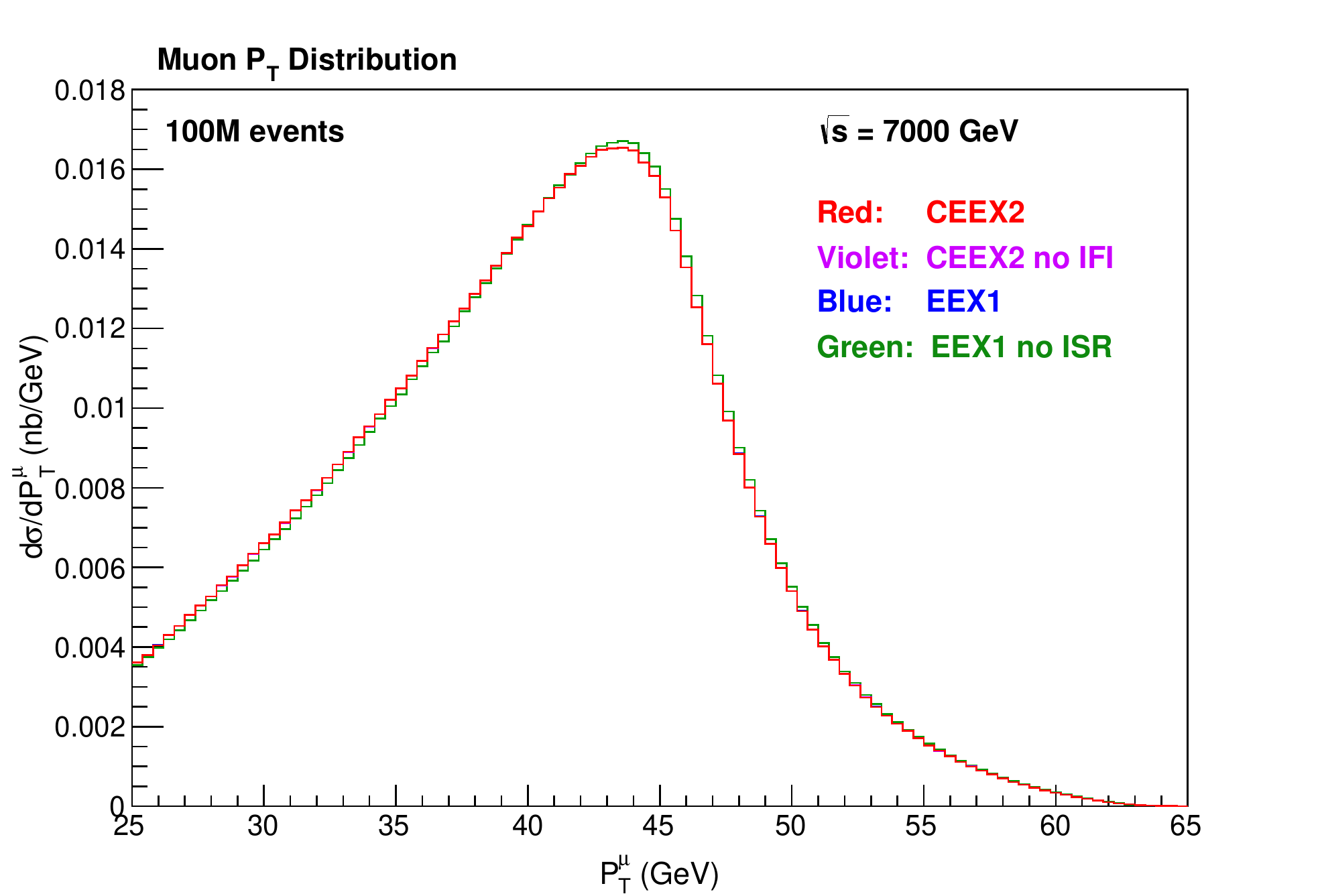}}
\put(3,0.2){\includegraphics[width=3in]{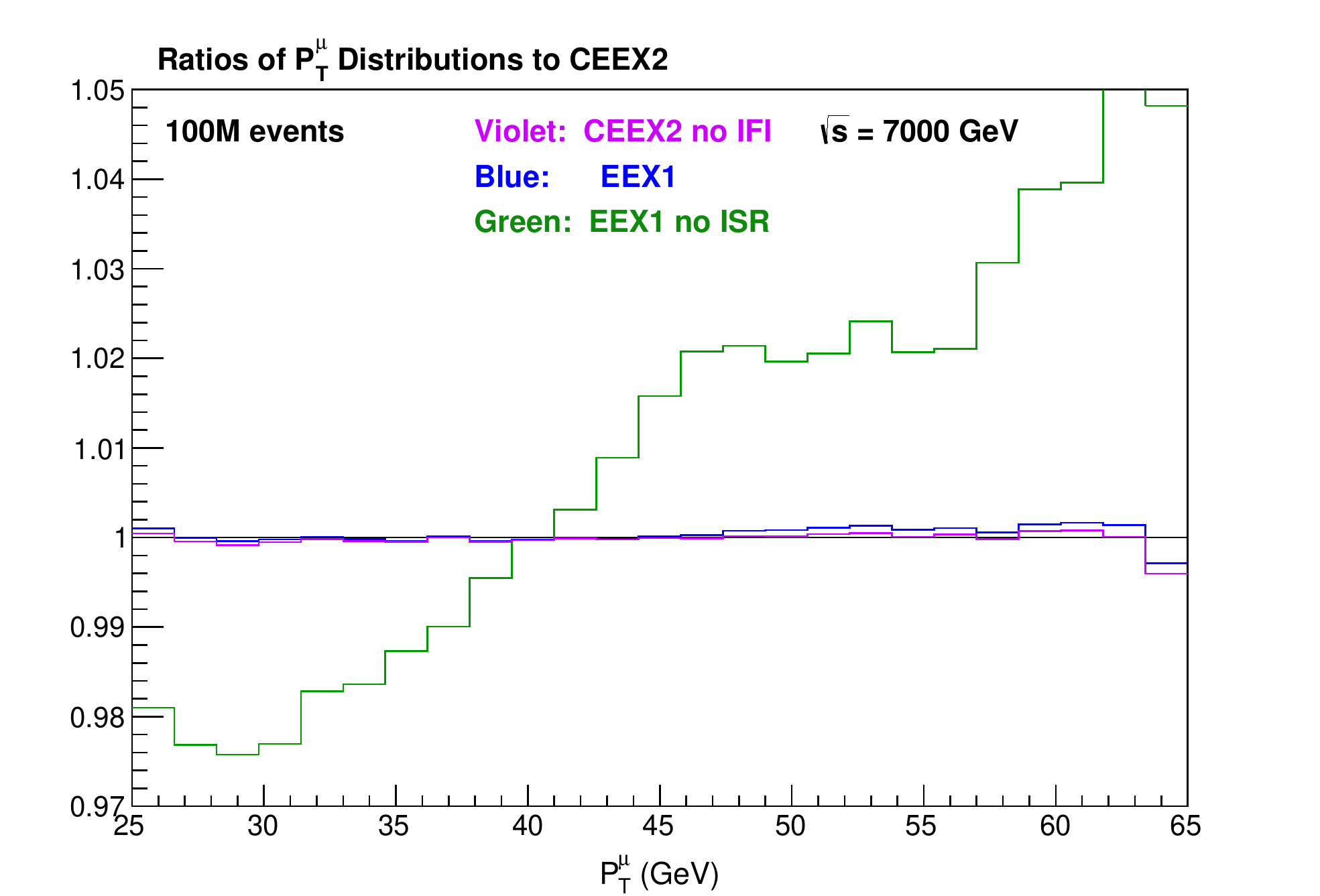}}
\end{picture}
\end{center}
\vspace{-10mm}
\caption{\baselineskip=11pt Muon transverse momentum distributions and their ratios for {\KK}MC-hh with the cuts specified in the text for the EW-CORR (electroweak-correction) labels ``CEEX2" (red -- medium dark shade), ``CEEX2 (no IFI)" (violet -- light dark shade), ``EEX1'' (blue -- dark shade), and ``EEX1 (no IFI)'' (green -- light shade), showered by HERWIG 6.5. The labels are explained in the text. The ratio plot features ``CEEX2'' as the reference distribution as noted in the respective title.}
\label{fig2}
\end{figure}

 
The ATLAS cuts used for assaying systematics in observables in W production and lepton pair decay events from the analogous observables in Z production and lepton pair decay events are:
$80 \text{GeV} < M_{\ell\bar{\ell}} < 100 \text{GeV},\; P^{\ell\bar{\ell}}_T < 30 \text{GeV}, \;\text{with both leptons satisfying} \; P^{\ell}_T >  25 \text{GeV}\; \text{and}\; |\eta_\ell| < 2.4$ . Here, for $\ell=e,\;\mu,$~ $M_{\ell\bar{\ell}}\; \text{and}\;  P^{\ell\bar{\ell}}_T$ are the lepton pair mass and transverse momentum, respectively, and $P^{\ell}_T\;\text{and}\;\eta_\ell$ are the lepton transverse momentum and pseudorapidity, respectively. For a detailed discussion of all observable distributions predicted by \KK{MC}-hh for these cuts, we refer to Ref.~\cite{kkmchh2}. To illustrate the type of effects we observe, we focus on the ATLAS data for $P^{\ell}_T$ reproduced here in Fig.~\ref{fig1} versus our \KK{MC}-hh 
prediction in Fig.~\ref{fig2}.
In Fig.~\ref{fig2}, the best prediction of the exact ${\cal O}(\alpha^2 L)$ CEEX calculation (labeled CEEX2) is compared with less precise predictions all of which we denote as follows:
\begin{itemize}
\item   $ {\cal O}(\alpha^2 L)$ CEEX with ISR+FSR+IFI -- labeled as ``CEEX2''\\
\item   ${\cal O}(\alpha^2 L)$ CEEX without IFI (initial state final state interference) -- labeled `` CEEX2 (no IFI)''\\
\item   ${\cal O}(\alpha)$ EEX -- labeled ``EEX1''\\
\item   ${\cal O}(\alpha)$ EEX without ISR (initial state radiation) -- labeled ``EEX1 (no ISR)''.\\
\end{itemize}
For completeness, to quantify the relative normalizations of these levels of precision we show their cross sections in Table 1:\\
\vbox{
\begin{center}
\begin{tabular}{|l|c|c|c|c|}
\hline
	    	& uncut (pb) & Difference	& cut (pb) & Difference\\
\hline
CEEX2 	    	& 844.74     	& $\times$  & 280.36 & $\times$\\
CEEX2 (no IFI) 	& 844.97 	& $+0.03\%$ & 280.31 & $-0.02\%$\\
EEX1  		& 844.45 	& $-0.03\%$ & 280.38 & $+0.007\%$\\
EEX1 (no ISR)  	& 844.97 	& $+0.03\%$ & 280.64 & $+0.10\%$\\
\hline
\end{tabular}
\\[1em]
{{\bf Table 1.} Total Cross Sections With and Without ATLAS cuts. Differences
are shown relative to CEEX2.}
\end{center}
}
There is a per mille level difference of the no-ISR cut cross section relative to the CEEX2 result; all other differences are at the fractional per mille level. The non-flat few \% level effects we see in Fig.~\ref{fig2} in the no-ISR differential spectrum are no where evident from the cross section normalizations. The IFI and the exact $ {\cal O}(\alpha^2 L)$ effects are below and at or below the per mille level, respectively.
We observe that the no-ISR curve from \KK{MC}-hh predicts the trends in the ATLAS data to be about 1-2\% higher than the theory ATLAS uses at the low end and a similar amount lower than the data at the high end. For, the theory ATLAS uses has only the QED ISR shower radiation which is not expected to reproduce effects due to transverse degrees of freedom very accurately.\par
\begin{figure}[ht]
\begin{center}
\includegraphics[width=160mm]{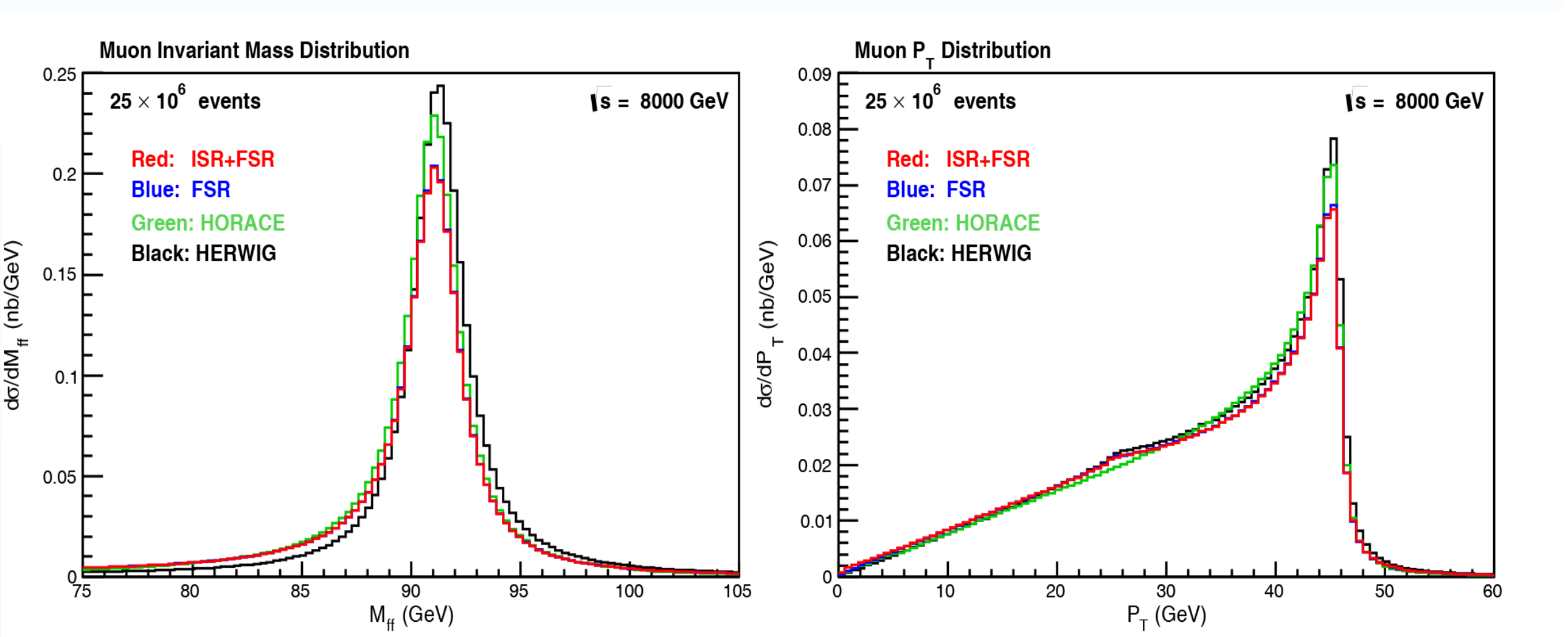}
\end{center}
\vspace{-4mm}
\caption{\baselineskip=11pt The muon pair and muon $P_T$ spectrum in {\KK}MC-hh and HORACE with the cuts specified in the text. The red(medium dark shade) curve corresponds to the EW-CORR switch CEEX ISR+FSR+EWK (ISR+FSR) for {\KK}MC-hh, the blue (dark shade) curve corresponds to the switch CEEX FSR+EWK (FSR)for {\KK}MC-hh and the green(light shade) curve corresponds to the switch \Order{\alpha} QED Shower FSR for HORACE, as explained in the text. For reference, we also show the unshowered Herwig result in the black curve.}
\label{fig3}
\end{figure}

In Refs.~\cite{kkmchh,radcor17say}, we have made a series of comparisons with HORACE~\cite{Horace1,Horace2,Horace3} in order to make a first step toward making contact with the benchmark studies in Ref.~\cite{vicini-dorw}, with the understanding that there still needs to be a proper tuning of the two calculations before such comparisons can be considered final. Here, we illustrate the type of results we find in Table 2, in which we compare the cross section normalizations, and in Fig.~\ref{fig3}, in which we show the muon pair invaraint mass and the muon transverse momentum spectra. In Table 2 and in Fig.~\ref{fig3}, we show comparisons for Z decays to muon pairs at 8 TeV with no shower in which we feature HORACE's best EW scheme with exponentiated FSR, so that it should agree with \KK{MC} with CEEX ${\cal O}(\alpha)$ exponentiation with ISR off, which we also feature. An unshowered HERWIG65~\cite{hwg} prediction without EW corrections is also shown for reference, as is a Born level HORACE result in Table 2. The cut on the generated $q\bar{q}$ invariant mass is 50 GeV $< M_{q\overline q} < $ 200 GeV. We also show the best \KK{MC}-hh CEEX ${\cal O}(\alpha^2 L)$ result in both Table 2 and in Fig.~\ref{fig3}.
 \vbox{
\begin{center}
\begin{tabular}{|l|c|c|c|}
\hline
MC	    	& EW Corrections 	& $\sigma$ (pb) & Difference\\
\hline
\KK{MC}-hh 	& CEEX2    	& $993\pm 1$   	& $\times$ \\
\KK{MC}-hh      & CEEX1 (no ISR)& $991\pm 1$ 	& $-0.20\%$\\
HORACE  	& ${\cal O}(\alpha)$ exp. & $1009.6\pm 0.4$ 	& $+1.7\%$\\
HORACE          & Born (no $\gamma$'s)    & $1025.2\pm 0.4$ 	& $+3.2\%$\\
HERWIG6.5       & Born (no $\gamma$'s)    & $1039.6\pm 0.2$     & $+4.7\%$\\
\hline
\end{tabular}
\\[1em]
{{\bf Table 2.} Total Cross Section Comparisons and Difference Relative to CEEX2}
\end{center}
}
In both the cross section normalizations in Table 2 and in the spectra in Fig.~\ref{fig3} we see differences at the level of 1.9\% and of 10\% respectively that we would expect to be significantly reduced when a proper tuning between the calculations is done.\par
To sum up, \KK{MC}-hh provides exact $ {\cal O}(\alpha^2 L)$ CEEX EW corrections to Z production and decay to lepton pairs for hadron-hadron scattering at high energies in the presence of a parton shower. The original exclusive YFS~\cite{yfs-1,yfs-2} exponentiation at the level of the squared amplitudes (EEX) for the QED radiation is also supported. We have illustrated new effects that should be taken into account for per mille level studies in precision LHC physics. We have presented a first step in comparison with other calculations using HORACE as an untuned vehicle and more work in this direction is in progress. We are also working on adding the exact QCD NLO correction following the methods in Refs.~\cite{mg5amcatnlo,krknlo} and on adding fermion pairs corrections.
One of us (BFLW) thanks Profs. S. Bethke and W. Hollik for the support and kind hospitality of Werner-Heisenberg-Institut, Max-Planck-Institut fuer Physik, Munich, Germany while part of this research was done. \par 
\par

\end{document}